# An observable effect of spin inertia in slow magneto-dynamics: Increase of the switching error rates in nanoscale ferromagnets[1]

**Rahnuma Rahman and Supriyo Bandyopadhyay**

Department of Electrical and Computer Engineering,
Virginia Commonwealth University, Richmond, VA 23284, USA

Corresponding author E-mail: sbandy@vcu.edu

The Landau-Lifshitz-Gilbert (LLG) equation, used to model magneto-dynamics in ferromagnets, tacitly assumes that the angular momentum associated with spin precession can relax instantaneously when the real or effective magnetic field causing the precession is turned off. This neglect of "spin inertia" is unphysical and would violate energy conservation. Recently, the LLG equation was modified to account for inertia effects. The consensus, however, seems to be that such effects would be unimportant in *slow* magneto-dynamics that take place over time scales much longer that the relaxation time of the angular momentum, which is typically few fs to perhaps ~100 ps in ferromagnets. Here, we show that there is at least one very serious and observable effect of spin inertia even in slow magneto-dynamics. It involves the switching error probability associated with flipping the magnetization of a nanoscale ferromagnet with an external agent, such as a magnetic field. The switching may take ~ns to complete when the field strength is close to the threshold value for switching, which is much longer than the angular momentum relaxation time, and yet the effect of spin inertia is felt in the switching error probability. This is because the ultimate fate of a switching trajectory, i.e. whether it results in success or failure, is influenced by what happens in the first few ps of the switching action when nutational dynamics due to spin inertia holds sway. Spin inertia *increases* the error probability, which makes the switching more error-prone. This has vital technological significance because it relates to the reliability of magnetic logic and memory.

---

# 1. Introduction

The Landau-Lifshitz-Gilbert (LLG) equation, widely used to model magneto-dynamics (damped precession) in ferromagnets, is known to have a shortcoming in that it ignores the effect of spin inertia that arises from spin-orbit coupling or other quantum effects [1-6]. The LLG equation is formulated on the assumption that two of the three principal moments of inertia associated with magnetization precession are zero and the third (corresponding to the kinetic energy of the precession) is non-zero. Gilbert, himself, famously opined: "I was unable to conceive of a physical object with an inertial tensor of this kind" [7]. Indeed, neglecting spin inertia would violate energy conservation if the damping rate is finite and this is why the traditional LLG equation is unphysical.

A modified LLG equation has now been derived to correct this error and it includes inertial effects in magneto-dynamics [8-16]. This new equation predicts the presence of *nutation* in magneto-dynamics. To understand "nutation", assume that the magnetic field (real or effective) causing the precession of the magnetization vector is switched off suddenly with zero damping. The precession should stop immediately because the field causing the Larmor precession has vanished. Without inertia, the magnetic moment also stops at this position, but if the kinetic energy is different from zero when the effective magnetic field vanishes, then the precession cannot be stopped abruptly and the magnetic moment must continue to rotate around the angular momentum vector in order to conserve energy [8]. In other words, precession is transformed into nutation, which is a consequence of spin inertia [8].

The nutation, however, lasts for a short while, typically a few fs to perhaps 100 ps in ferromagnets, after which the angular momentum equilibrates. Therefore, it is natural to assume that spin inertia would not have a lasting effect in slow phenomena that take ~ns to complete. In this paper, we examine this conjecture and show that it is not correct; there is at least one situation where it is untrue. The case in point involves the "switching error rate" associated with switching the magnetization of a bistable nanomagnet from one stable state to the other with an external agent. The switching takes ~ns to complete. Hence, one would expect that spin inertia would not have much of an effect on the error rate. However, it turns out that spin inertia increases the switching error probability considerably, especially when the strength of the switching field is close to the threshold value for switching. This happens because the fate of a switching trajectory, i. e. whether it ends up being a switching success or failure, is influenced by what happens in the first few ps after the switching field is turned on (i. e. when nutation takes place), even though the entire switching process takes ~ns. This is reminiscent of "chaos theory", where small differences in initial conditions can yield very different outcomes even if the system is deterministic, meaning that it evolves according to a deterministic equation [17-19]. Stated differently, the system trajectory is extremely sensitive to initial condition or the initial behaviour. In our case, the initial behaviour is governed by nutation arising from spin inertia. Consequently, spin inertia has an influence on the outcome of switching that takes ~ns to complete. Thermal noise introduces an additional stochasticity in the system and that can exacerbate this effect.

Such an observation has important technological significance. Nanomagnetic logic and memory devices carry out digital information processing by flipping the magnetization of a bistable nanomagnet from one stable state to another with an external agent such as a magnetic field, spin-polarized current, strain, etc. To be able to perform the flipping action *reliably* is always a critical requirement. If spin inertia exacerbates switching failure, then that would impact the reliability of magnet-based digital processors and storage elements and may even impugn the viability of nanomagnetic logic [20].

In the past, there have been a number of theoretical studies to calculate the switching error probabilities in magnetization reversal actuated by low-energy mechanisms such as strain [21-29] and voltage controlled magnetic anisotropy (VCMA) [30, 31]. Almost all of them have used the stochastic Landau-Lifshitz-Gilbert (s-LLG) equation [sometimes also referred to as the Landau-Lifshitz-Gilbert-Langevin equation] to model the time evolution of the magnetization of a bistable nanomagnet perturbed by an effective switching magnetic field representing strain or VCMA in the presence of thermal noise to estimate the error probability. None of them considered spin inertia. Here, we have revisited this problem and calculated switching error probabilities taking spin inertia into account. We found that the error rates increase when spin inertia is taken into account and the increase is significant when the switching field is close to the threshold value. Furthermore, the longer it takes for the angular momentum to equilibrate (i. e. the longer the nutation lasts), the more is the error probability. Therefore, it has important implications for *low-energy* switching which would typically be carried out with slightly sub-threshold switching fields to reduce energy dissipation.

In Section 2, we formulate the simulation strategy for modelling magnetization switching dynamics in the presence of spin inertia and in section 3, we present results. Finally, in section 4, we present the conclusions.

## 2. Simulation Approach

The modified LLG equation that accounts for spin inertia is [1]

$$\frac{d\vec{m}(t)}{dt} = -|\gamma|\vec{m}(t) \times \left[\vec{H}_{eff}(t) - \frac{\alpha}{|\gamma|}\left(\frac{d\vec{m}(t)}{dt} + \tau\frac{d^2\vec{m}(t)}{dt^2}\right)\right], \quad (1)$$

where $\vec{m}(t)$ is the magnetization vector normalized to the saturation magnetization $M_s$, $\vec{H}_{eff}(t)$ is the effective or real magnetic field applied to make the magnetization switch and around which the magnetization vector precesses before equilibrating, $\gamma$ is the gyromagnetic precession constant, $\alpha$ is the Gilbert damping factor associated with damping of the precession, and $\tau$ is the relaxation time of the angular momentum (the timescale over which inertial effects are thought to be important). Generally, $\tau = I_1/(\alpha M_s^2) = I_2/(\alpha M_s^2)$, where $I_1$ and $I_2$ ($I_1=I_2$) are the principal moments of inertia components in the two directions orthogonal to the magnetization vector. The value of $\tau$ can range from a few fs to ~100 ps in ferromagnets [15]. The usual LLG equation assumes $I_1 = I_2 = 0$ and hence $\tau = 0$. The only difference between the usual LLG equation and the modified one is the second derivative term in the right-hand-side of Equation (1).

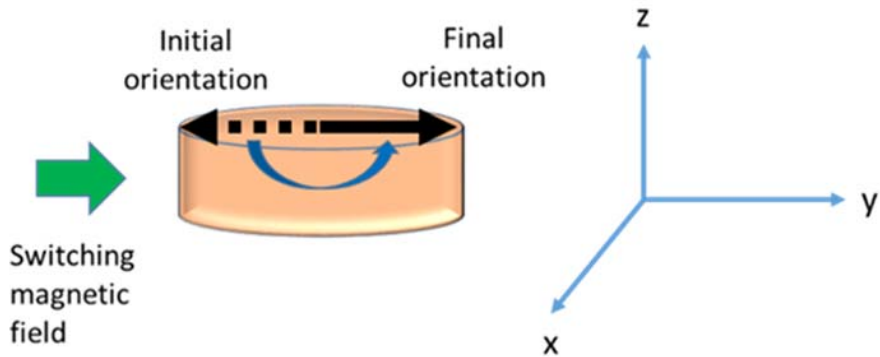

**Figure 1:** (a) Magnetization reversal in a single domain nanomagnet shaped like an elliptical disk due to an external switching magnetic field directed along the major (easy) axis.

To study what effect spin inertia might have on magnetic reversal phenomena, we simulate the flipping of the magnetization of a nanomagnet shaped like an elliptical disk (as shown in figure 1) when subjected to a magnetic field directed opposite to the initial magnetization orientation. The initial magnetization orientation is in one direction along the major axis (easy axis) and magnetic reversal (flipping) would take it to the opposite direction along the major axis. The simulation is carried out for various values of τ to assess the importance of spin inertia in this slow magneto-dynamical phenomenon. Here, thermal noise has two effects: first, it causes some switching failures when the switching field is above the threshold value and second, it also allows switching to occur successfully with non-zero probability when the switching field is sub-threshold. Our simulations reveal that spin inertia increases the error probability (probability that switching is unsuccessful) at any switching field and the increase is most prominent when the switching field is close to the threshold value.

The nanomagnet parameters used in our simulations are given in Table 1. The Gilbert damping constant is representative of cobalt nanomagnets.

**Table 1: Nanomagnet parameters**

| Major axis (*a*) | 105 nm |
|---|---|
| Minor axis (*b*) | 95 nm |
| Thickness (*t*) | 6 nm |
| Damping constant (α) | 0.1 |
| Relaxation time (τ) | 10 ps |

Equation (1) can be written in Cartesian coordinates as

$$\frac{dm_x(t)}{dt} = -|\gamma|(m_y(t)H_z - m_z(t)H_y)$$
$$+\alpha\left(m_y(t)\frac{dm_z(t)}{dt} - m_z(t)\frac{dm_y(t)}{dt}\right)$$
$$+\alpha\tau\left(m_y(t)\frac{d^2m_z(t)}{dt^2} - m_z(t)\frac{d^2m_y(t)}{dt^2}\right)$$

$$\frac{dm_y(t)}{dt} = -|\gamma|(m_z(t)H_x - m_x(t)H_z)$$
$$+\alpha\left(m_z(t)\frac{dm_x(t)}{dt} - m_x(t)\frac{dm_z(t)}{dt}\right)$$
$$+\alpha\tau\left(m_z(t)\frac{d^2m_x(t)}{dt^2} - m_x(t)\frac{d^2m_z(t)}{dt^2}\right)$$

$$\frac{dm_z(t)}{dt} = -|\gamma|(m_x(t)H_y - m_y(t)H_x)$$
$$+\alpha\left(m_x(t)\frac{dm_y(t)}{dt} - m_y(t)\frac{dm_x(t)}{dt}\right)$$
$$+\alpha\tau\left(m_x(t)\frac{d^2m_y(t)}{dt^2} - m_y(t)\frac{d^2m_x(t)}{dt^2}\right) \quad , \quad (2)$$

where $m_i(t)$ is the $i$-the component of the normalized magnetization vector (normalized to saturation magnetization) and $H_i$ is the $i$-th component of the effective magnetic field that the magnetization experiences.

Equation (2) represents the modified LLG equation with spin inertia included [8, 10]. In the absence of any spin inertia effect, the traditional LLG equation would be Equation (2) without the terms involving the quantity $\tau$.

The effective magnetic field components are given by

$$H_x = -M_s N_{d-xx} m_x + \sqrt{\frac{2\alpha kT}{|\gamma|(1+\alpha^2)\mu_0 M_s \Omega \Delta t}} G_{(0,1)}^x(t)$$

$$H_y = -M_s N_{d-yy} m_y + \sqrt{\frac{2\alpha kT}{|\gamma|(1+\alpha^2)\mu_0 M_s \Omega \Delta t}} G_{(0,1)}^y(t) + H_{bias} \quad (3)$$

$$H_z = -M_s N_{d-zz} m_z + \sqrt{\frac{2\alpha kT}{|\gamma|(1+\alpha^2)\mu_0 M_s \Omega \Delta t}} G_{(0,1)}^z(t)$$

where the first term in the right-hand-side is the demagnetization field associated with shape anisotropy and the second term is the random magnetic field due to thermal noise [32]. Here, $\Omega$ is the volume of the nanomagnet ($\Omega = \frac{\pi}{4} \times a \times b \times t$), $\mu_0$ is the permeability of free space, $\Delta t$ is the time step used in the simulation (0.01 ps), $G_{(0,1)}^i(t)$ [$i = x, y, z$] are three uncorrelated Gaussians of zero mean and unit standard deviation, $H_{bias}$ is the switching magnetic field directed along the $+y$ axis as shown in figure 1, and the demagnetization parameters for the elliptical disk are [33]

$$N_{d-xx} = \frac{\pi}{4}\left(\frac{t}{a}\right)\left[1 + \frac{5}{4}\left(\frac{a-b}{a}\right) + \frac{21}{16}\left(\frac{a-b}{a}\right)^2\right]$$

$$N_{d-yy} = \frac{\pi}{4}\left(\frac{t}{a}\right)\left[1 - \frac{1}{4}\left(\frac{a-b}{a}\right) - \frac{3}{16}\left(\frac{a-b}{a}\right)^2\right] \quad (4)$$

$$N_{d-zz} = 1 - N_{d-xx} - N_{d-yy}$$

We assume that the temperature is room temperature ($T$ = 300 K) in all our simulations.

We can define a "threshold" magnetic field for switching by equating the magneto-static energy due to the applied field to twice the shape anisotropy energy barrier in the plane of the nanomagnet. This yields an expression for the threshold field $H_{thres}$:

$$\mu_0 M_s^2 \Omega (N_{d-xx} - N_{d-yy}) = \mu_0 M_s H_{thresh} \Omega$$
$$\Rightarrow H_{thresh} = M_s (N_{d-xx} - N_{d-yy}) \quad (5)$$

For the parameters given in Table 1, $N_{d-xx} = 0.050$; $N_{d-yy} = 0.044$; $N_{d-zz} = 0.906$ and $H_{thresh}$ = 6000 A/m. Note that in the presence of thermal noise, it is possible to switch magnetization even with a sub-threshold magnetic field since thermal fluctuations can aid the magnetization to overcome the depressed shape anisotropy energy barrier (depressed by the magnetic field) and cause magnetization reversal.

In our simulations, the switching field is turned on abruptly at time $t = 0$. We use the initial conditions

$$m_x(0) = 0.1$$
$$m_y(0) = -0.99 \quad (6)$$
$$m_z(0) = 0.1$$

and continue to simulate the dynamics with Equation (2) and also the traditional LLG equation until $m_y$ approaches and lingers around either +1 or -1. The former indicates switching success and the latter switching failure. To compute switching error probability, we solve Equations (2) and the traditional LLG equation for 1000 switching trajectories at each value of $H_{bias}$. The switching error probability is the fraction of trajectories that end in failure. We then compare the results from Equations (2) [spin inertia included] and the traditional LLG [spin inertia excluded] to see if spin inertia makes any difference.

## 3. Results and discussion

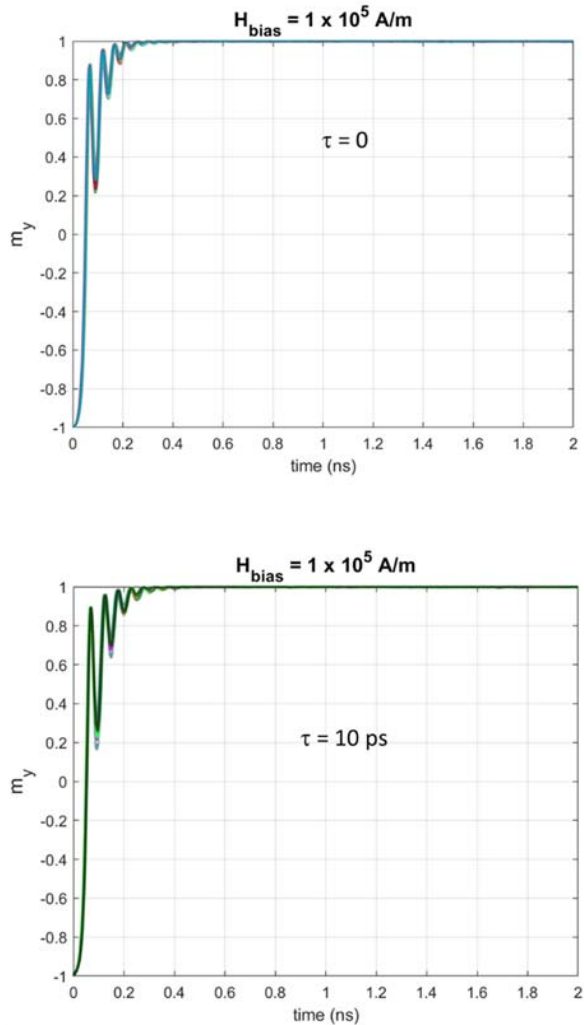

**Figure 2:** Time evolution of the magnetization component $m_y$ in the direction of the switching field when the switching field is well above threshold. The upper panel shows the result without spin inertia and the lower panel the result with spin inertia.

In figure 2, we show the time evolution of the magnetization component along the major axis (easy axis) of the elliptical nanomagnet (i. e. $m_y(t)$) after a switching field of $10^5$ A/m directed along the +y axis is switched on abruptly at $t = 0$. We show the time evolution of 25 arbitrarily picked trajectories, which are practically identical to each other (very little difference between them) since the switching field is well above the threshold switching field in this case. The top panel in figure 2 shows the result without spin-inertia and the bottom one the result with spin inertia included where the relaxation time $\tau$ has been taken to be 10 ps. There is almost no perceptible difference between the two plots when the switching field is well above the threshold field. In other words, spin inertia does not have much of an effect when the switching field is well above threshold.

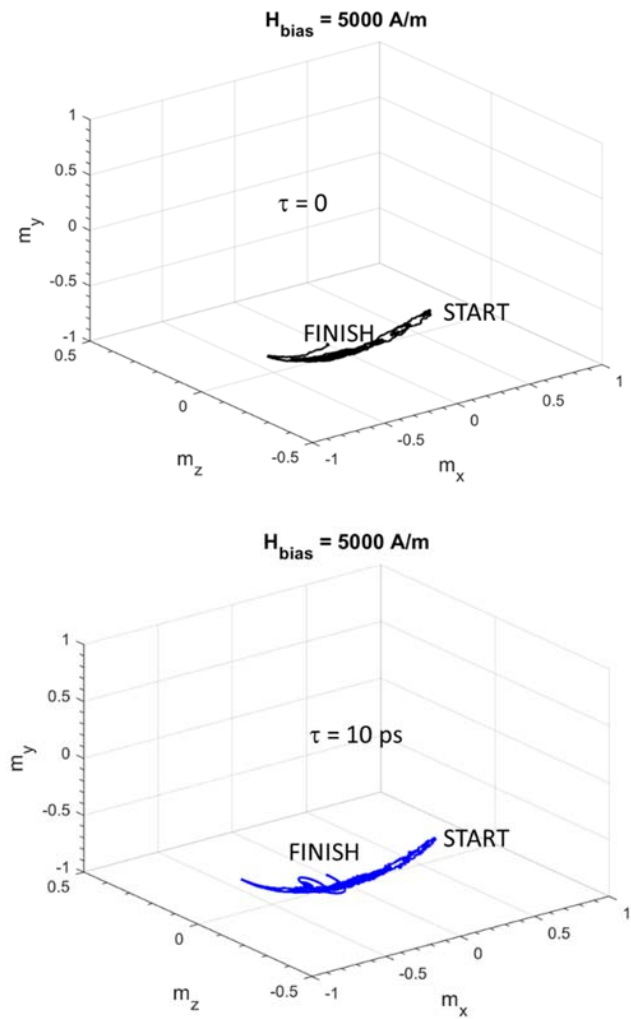

**Figure 3:** An arbitrarily picked switching trajectory (that failed to switch) under a slightly sub-threshold switching field of 5000 A/m at room temperature. The upper panel shows the result in the absence of spin inertia and the lower panel shows the result in the presence of spin inertia ($\tau = 10$ ps). The simulation was carried out for 5 ns in both cases.

In figure 3, we show an arbitrarily picked switching trajectory at room temperature under a switching field of 5000 A/m (close to the threshold field, slightly sub-threshold), with the upper panel showing the result without spin inertia and the lower panel showing the result with spin inertia ($\tau = 10$ ps). This trajectory resulted in switching failure in both cases ($m_y$ went from ~ -1 to ~ -1 and did not change to +1).

Two interesting features are observed in figure 3. First, there are obviously more exaggerated precessions around the termination point in the presence of spin inertia, which must be related to nutation dynamics [15] since that is the only difference between the two cases. This is a striking example showing that the nutation which takes place at the beginning of the trajectory influences what happens at the end.

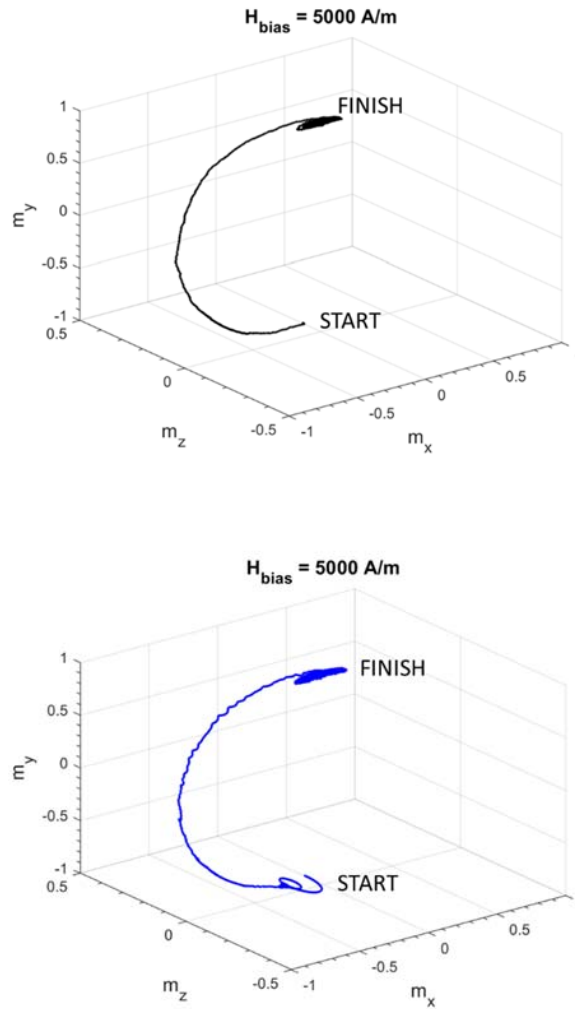

**Figure 4:** Another arbitrarily picked switching trajectory under a slightly sub-threshold switching field of 5000 A/m at room temperature. The upper panel shows the result in the absence of spin inertia and the lower panel shows the result in the presence of spin inertia ($\tau = 10$ ps). The simulation was carried out for 5 ns in both cases. In this case, switching succeeded in both cases.

In figure 4, we show another arbitrarily picked trajectory, but in this case, the switching succeeded ($m_y$ went from ~ -1 to ~ +1). The effect of nutation is very visible in this trajectory where we see wiggles around the starting point. These are clear signatures of nutation.

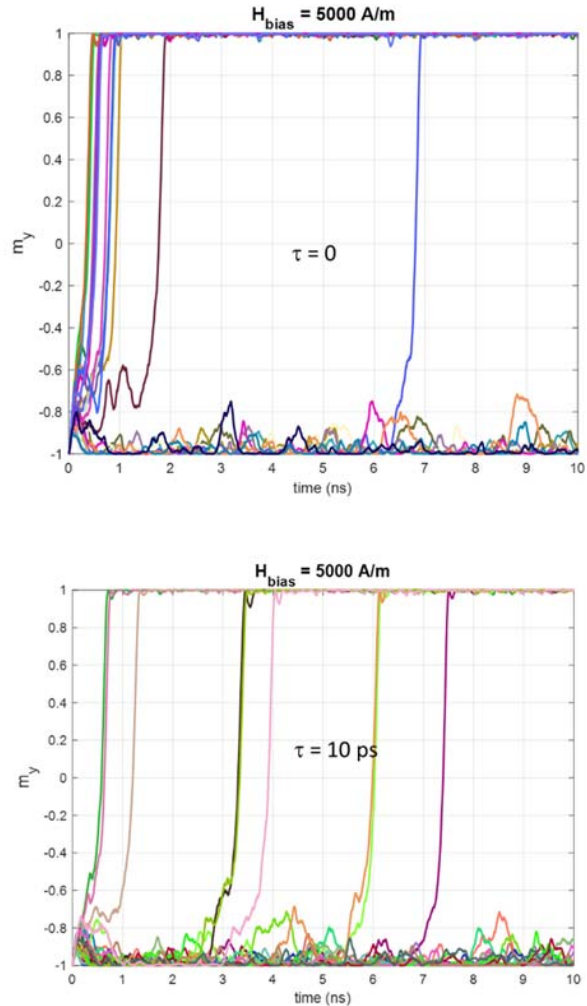

**Figure 5:** Time evolution of the magnetization vector's component along a (slightly sub-threshold) switching magnetic field of 5000 A/m in the case of 25 arbitrarily picked trajectories in the presence of room temperature thermal noise. 9 out of 25 trajectories fail to switch in the absence of spin inertia, while 15 fail to switch in the presence of spin inertia, ($\tau$ = 10 ps) as shown in the lower panel.

In figure 5, we show the time evolution of the magnetization component along the major (easy) axis of the magnetization (which is also the direction of the switching field) for 25 arbitrarily picked trajectories when a slightly sub-threshold field of 5000 A/m is used to swutch. In the absence of spin inertia, 9 out of 25 trajectories failed to switch, while in the presence of spin inertia ($\tau$ = 10 ps), 15 out of 25 trajectories failed. In the former case, the switching error probability was 36%, while in the latter case, it increased to 60%. This is a very non-trivial consequence of spin inertia. It shows up in a slow phenomenon that takes ~ns to complete, where one would not have expected spin inertia to have much of an effect, let alone such a non-trivial consequence. Here, the time scale for switching is 100 times larger than the value of $\tau$ which was taken to be 10 ps for this simulation. *Hence, spin inertia can have serious consequences in even slow dynamics.* Additionally, this result is of immense practical importance since the switching error probability is a measure of the reliability of magnetic switches widely used in both magnetic logic and memory.

Finally, in figure 6, we show the switching error probability as a function of the switching magnetic field, calculated by simulating 1000 trajectories in the presence of thermal noise for each data point. The simulation time for each trajectory was 25 ns, by which time every trajectory either resulted in switching success $(m_y \approx +1)$ or failure $(m_y \approx -1)$.

Note from figure 6 that at the slightly sub-threshold switching field of 5000 A/m, spin inertia increased the switching error probability from 40% to 70% when $\tau = 100$ ps. The switching error probability increases if $\tau$ is increased because a larger value of $\tau$ corresponds to a longer duration of nutation dynamics.

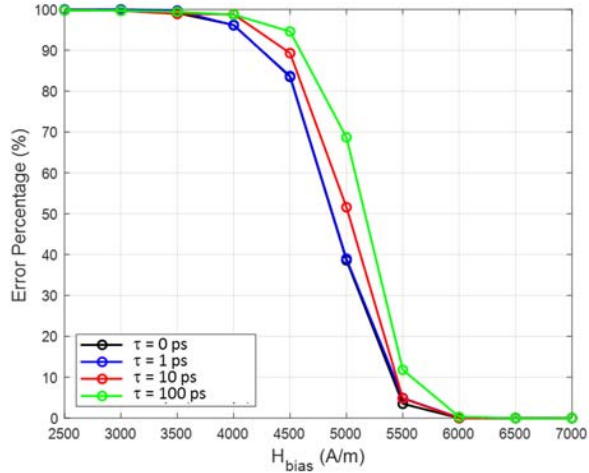

**Figure 6:** Switching error probability versus switching magnetic field strength in the presence of room temperature thermal noise for $\tau = 0$ ps, 1 ps, 10 ps and 100 ps.

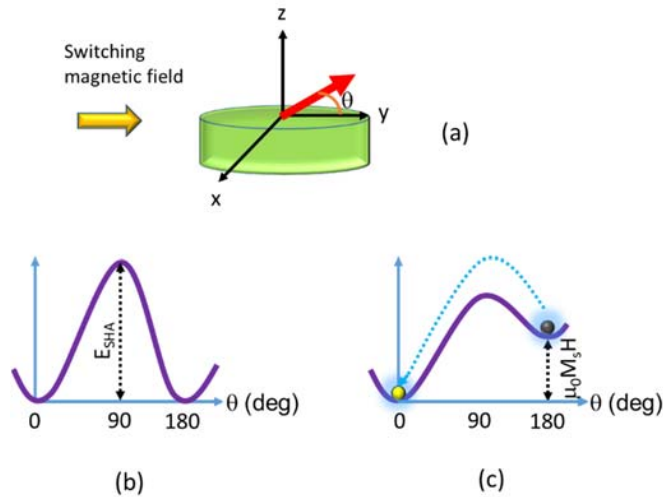

**Figure 7:** (a) The magnetization vector subtends an angle θ with the right segment of the major axis of the elliptical nanomagnet. (b) The potential profile in the plane of the nanomagnet as a function of the magnetization orientation θ in the absence of any magnetic field, where $E_{SHA}$ is the energy barrier due to shape anisotropy. (c) The same potential profile in the presence of a slightly sub-threshold magnetic field directed along the major axis and pointing to the right. The grey "ball" denotes the initial state of the nanomagnet and the yellow ball the final state if a successful switching event takes place.

To understand how spin inertia can increase the switching failure rate, consider the potential energy profile in the plane of the nanomagnet as a function of the magnetization orientation, shown schematically in figures 7(b) and 7(c) in the absence and presence of an applied switching magnetic field, respectively. The two minima represent the two stable magnetization orientations along the major (easy) axis of the elliptical nanomagnet and the energy barrier separating them is due to shape anisotropy. The magnetic field depresses the energy barrier as shown in figure 7(c), in the process lifting the degeneracy of the two minima. A sub-threshold magnetic field will leave some energy barrier still separating the two minima. Any residual angular momentum of the precession associated with nutation dynamics may hinder the magnetization from overcoming the vestigial energy barrier and thus tend to prevent magnetization reversal (switching). This is why spin inertia can increase the switching failure rate.

4. **Conclusion**

In this work, we have shown that spin inertia, which has hitherto been neglected in almost all calculations of switching error probability associated with magnetic reversal in nanomagnets, has a surprising effect on the switching error probability. The effect is strongest when the switching field is close to the threshold value. One would of course prefer to switch with sub-threshold switching fields to minimize the energy dissipation that accompanies switching and hence this is an important result. Spin inertia increases switching error rates in low-energy switching.

We also mention that whether spin inertia increases or decreases error rate may be case-specific. It is entirely possible that there may be some pathological cases where spin inertia might suppress the error probability. We have not searched for such cases since our primary objective was to demonstrate that, contrary to popular belief, the effect of spin inertia is not just restricted to short time scales over which the angular momentum equilibrates, which is a few ps, but it also leaves its signature on phenomena occurring over much longer time scales (~ns). Thus, to observe the effect of spin inertia, one does not necessarily require THz dynamics and even GHz dynamics may suffice. This sheds new light on our understanding of the implications of spin inertia.

**Acknowledgement**: The authors acknowledge support from the US National Science Foundation under grant CCF-2001255.